%% file: main.tex
\documentclass[conference]{IEEEtran}

\usepackage [norelnum,twocolumn] {figures}

\usepackage{cite}
\usepackage{amsmath,amssymb,amsfonts}
\usepackage{algorithmic}
\usepackage{graphicx}
\usepackage{textcomp}
\usepackage{xcolor}
\usepackage{csquotes}
\usepackage{booktabs}
\def\BibTeX{{\rm B\kern-.05em{\sc i\kern-.025em b}\kern-.08em
    T\kern-.1667em\lower.7ex\hbox{E}\kern-.125emX}}

\usepackage{orcidlink}
\hypersetup{pdfborder={0 0 0}}

\usepackage[nolist]{acronym}
\input{acronyms}

\usepackage[noabbrev,capitalise]{cleveref}

\usepackage{siunitx}
\usepackage{eurosym}
\DeclareSIUnit{\sieuro}{\mbox{\euro}}

\usepackage[inline]{enumitem}

\usepackage[htt]{hyphenat}

\usepackage{silence}
\makeatletter
\newcommand{\linebreakand}{%
  \end{@IEEEauthorhalign}
  \hfill\mbox{}\par
  \mbox{}\hfill\begin{@IEEEauthorhalign}
}
\makeatother

\title{Automatic Generation of Combinatorial Reoptimisation Problem Specifications: A Vision}

\author{%
    \IEEEauthorblockN{Maximilian Kratz
    \orcidlink{0000-0001-7396-7763}}
    \IEEEauthorblockA{\textit{Real-Time Systems Lab} \\
    \textit{Technical University of Darmstadt}\\
    Darmstadt, Germany \\
    maximilian.kratz@es.tu-darmstadt.de}
\and
    \IEEEauthorblockN{Steffen Zschaler
    \orcidlink{0000-0001-9062-6637}}
    \IEEEauthorblockA{\textit{Department of Informatics} \\
    \textit{King's College London}\\
    London, United Kingdom \\
    szschaler@acm.org}
\and
    \IEEEauthorblockN{Jens Kosiol
    \orcidlink{0000-0003-4733-2777}}
    \IEEEauthorblockA{\textit{Software Engineering Group} \\
    \textit{Philipps-Universität Marburg}\\
    Marburg, Germany \\
    kosiolje@mathematik.uni-marburg.de}
\linebreakand
    \IEEEauthorblockN{Gabriele Taentzer
    \orcidlink{0000-0002-3975-5238}}
    \IEEEauthorblockA{\textit{Software Engineering Group} \\
    \textit{Philipps-Universität Marburg}\\
    Marburg, Germany \\
    taentzer@mathematik.uni-marburg.de}
}

\begin{document}
    \maketitle

    \begin{abstract}
		Once an optimisation problem has been solved, the solution may need adaptation when contextual factors change.
		This challenge, also known as \emph{reoptimisation}, has been addressed in various problem domains, such as railway crew rescheduling, nurse rerostering, or aircraft recovery.
		This requires a modified problem to be solved again to ensure that the adapted solution is optimal in the new context.
		However, the new optimisation problem differs notably from the original problem:
		\begin{enumerate*}[label={(\roman*)}, itemjoin={;\ }, itemjoin*={; and\ }]
			\item we want to make only minimal changes to the original solution to minimise the impact
			\item we may be unable to change some parts of the original solution (e.g., because they refer to past allocations)
			\item we need to derive a change script from the original solution to the new solution.
		\end{enumerate*}
		In this paper, we argue that Model-Driven Engineering (MDE)---in particular, the use of declarative modelling languages and model transformations for the high-level specification of optimisation problems---offers new opportunities for the systematic derivation of reoptimisation problems from the original optimisation problem specification.
		We focus on combinatorial reoptimisation problems and provide an initial categorisation of changing problems and strategies for deriving the corresponding reoptimisation specifications.
		We introduce an initial proof-of-concept implementation based on the GIPS (Graph-Based (Mixed) Integer Linear Programming Problem Specification) tool and apply it to an example resource-allocation problem: the allocation of teaching assistants to teaching sessions.
    \end{abstract}

    \acresetall


    \input{introduction}
    \input{motivating_example}
    \input{problem_types}
    \input{strategies}
    \input{implementation}
    \input{related_work}
    \input{roadmap}

   \section*{Acknowledgments}
    The authors would like to thank Andy Schürr for his valuable input and insightful discussions, which contributed to the development of this work.

    This work was partially funded by the German Research Foundation (DFG), project \enquote{Model-Driven Optimization in Software Engineering} (\href{https://gepris.dfg.de/gepris/projekt/462887453?language=en}{TA 294/19-1}) and by Short-Term Scientific Missions \href{https://www.roar-net.eu/grants/stsm/cd9e1cb8/}{\enquote{Towards Deriving Incremental Optimization Problems from Batch Specifications of ROAs}} and \href{https://www.roar-net.eu/grants/stsm/16fb8b8f/}{\enquote{From Models to the ROAR-NET API and Back}} in the context of the \textit{COST Action Randomised Optimisation Algorithms Research Network} (ROAR-NET, \url{https://www.roar-net.eu}), CA22137, supported by COST (European Cooperation in Science and Technology).

    \bibliographystyle{IEEEtranDOI}
    \bibliography{literature}
\end{document}

%% file: acronyms.tex
\begin{acronym}
    \acro{CSP}{Constraint Satisfaction Problem}
    \acro{DSL}{Domain-Specific Language}
    \acro{EMF}{Eclipse Modeling Framework}
    \acro{FRQ}{Future Research Question}
    \acro{GIPS}{\textbf{G}raph-Based (M)\textbf{I}LP \textbf{P}roblem \textbf{S}pecification}
    \acro{GIPSL}{\textbf{G}raph-Based (M)\textbf{I}LP \textbf{P}roblem \textbf{S}pecification \textbf{L}anguage}
    \acro{GT}[GT]{Graph Transformation}
    \acro{IDE}{Integrated Development Environment}
    \acro{IGPM}{Incremental Graph Pattern Matching}
    \acro{ILP}[ILP]{Integer Linear Programming}
    \acro{LHS}{Left-Hand Side}
    \acro{MDE}{Model-Driven Engineering}
    \acro{MDSE}[MDSE]{Model-Driven Software Engineering}
    \acro{MDO}{Model-Driven Optimisation}
    \acro{MILP}{Mixed-Integer Linear Programming}
    \acro{MT}{Model Transformation}
    \acro{OCL}{Object Constraint Language}
    \acro{PM}{Pattern Matching}
    \acro{RQ}{Research Question}
    \acro{RHS}{Right-Hand Side}
    \acro{TA}{Teaching Assistant}
    \acro{UML}{Unified Modeling Language}
\end{acronym}

%% file: introduction.tex

\section{Introduction}
\label{sec:introduction}

    Many problems require solutions that are optimal in some way.
    Techniques for specifying and solving optimisation problems have been studied for a long time and optimisation has been applied in many different domains~\cite{MN21}.
    Optimisation problem specifications consist of three key components:
    \begin{enumerate}
        \item a specification of what aspects of the problem can be controlled---traditionally captured through so-called \emph{decision variables} whose values can be varied by different candidate solutions;
        \item a specification of the \emph{search space}---often captured through \emph{constraints} over the values of decision variables and optional helper variables that must be satisfied by each feasible candidate solution; and
        \item a mechanism for ranking the quality of different feasible solutions---often captured through one or more \emph{objective functions} assigning numerical values to feasible solutions.
    \end{enumerate}

    Depending on the details of these components, many different types of optimisation problems can be differentiated.
    Here, we focus on (potentially multi-objective) combinatorial optimisation problems~\cite{PS98}; that is, problems where decision variables can only hold values from discrete sets of elements.

    \insertFigure[targetwidth=1\linewidth,%
          caption={Overview of the proposed approach to systematically derive reoptimisation problems and their solution strategies. The bottom half of the figure shows an activity diagram of our approach, containing the solver runs (problem instance as input fed into an off-the-shelf solver resulting in a corresponding solution). The top half shows a class diagram of our contribution on how to derive reoptimisation problems from a given optimisation problem and original solver specification.}%
          ]%
         {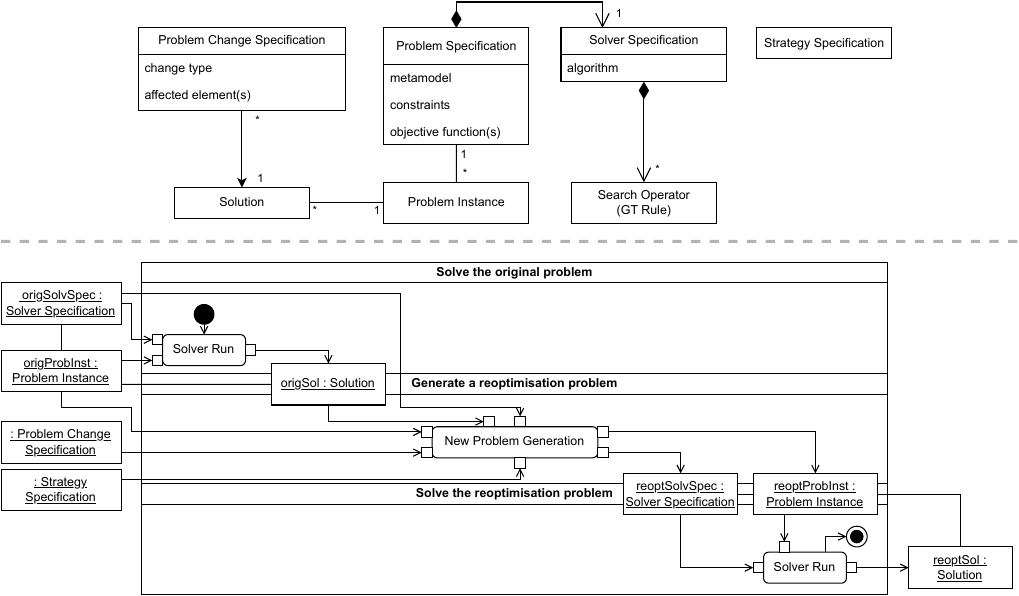}

    In many application scenarios, the optimality of a solution depends on contextual conditions.
    Solutions will not remain optimal forever, but will need to be adapted as conditions change.
    Such an adaptation requires the problem to be reoptimised, which involves solving two optimisation problems~\cite{SST+18}:
    (1) computing an optimal (or close to optimal) solution for the new problem instance, and (2) efficiently converting the current solution to the new one.
    This general setting has been discussed under different names for different concrete optimisation problems, such as railway crew rescheduling~\cite{VPHK12}, nurse rerostering~\cite{MV11}, and aircraft recovery~\cite{SVMP23}.
    More encompassing views also exist---under different titles, and sometimes with slight variations in what exactly is considered---such as reoptimisation~\cite{SST+18}, minimal perturbation problems~\cite{Bartak+04}, dynamic optimisation~\cite{NYB12}, and incremental optimisation~\cite{COG+18}.
    \emph{However, a general easy-to-use framework for systematically---ideally automatically---deriving reoptimisation specifications from an original problem specification, an original optimal solution, and a set of changes does not exist.}
    Instead, cumbersome manual respecification seems to be the state of the art.

    Techniques from \ac{MDE} have been used to enable users with limited technology knowledge to specify multi-objective combinatorial optimisation problems and their solution strategies~\cite{Fleck+15,Burdusel+18,John+19,John+23,gipsGCM2022}.
    The search space is typically defined by a modelling language, often specified declaratively through metamodels and \ac{OCL} constraints. Controllable problem parts are specified through \ac{GT} rules, and fitness functions are provided as model queries.
    A concrete instance of the optimisation problem is given by an instance model of the metamodel, as are candidate solutions.
    Optimisation problems are then solved using evolutionary techniques (e.g.,~\cite{John+19,Burdusel+18,Fleck+15}) or using solvers for \ac{MILP} problems (e.g.,~\cite{gipsGCM2022}).

    {\em As \ac{MDE} enables users to specify both the optimisation problem and the solver strategies using explicit, declarative models, new opportunities arise for conceptualising the relationship between the original and reoptimisation problems, as well as their respective solver strategies.
    This clear conceptual relationship will form the basis for developing a generator of reoptimisation problems and solver specifications from the original optimisation problem and solver specifications, eliminating the need for manual respecification.}

    \Cref{figure:overview} shows an overview diagram of the approach we are proposing in this paper.
    Starting from an \emph{original problem specification}
    ---consisting of a metamodel, constraints, objective functions, and search operators (typically a set of transformation rules)---it is possible to compute an initial solution given a specific problem instance.
    At a later point in time, changes may occur that make this solution suboptimal or even invalid.
    Different types of changes can occur and they may require different adjustments to the problem specification so that a new solution can be found.
    We propose that different types of changes could be classified in \emph{problem change type specifications,} which can be instantiated to describe the specific change that has occurred (we call this a \emph{specific change description} in the figure).
    The problem instance, original problem specification, and specific change description together can be used to generate a \emph{reoptimisation problem specification}, which can be solved by a standard solver again to create a new solution that is valid and optimal for the new circumstances.
    This \emph{new problem generation} process is the focus of this paper.
    We will discuss different strategies that can be employed in this process---\Cref{figure:overview} indicates this by parametrising the process with a \emph{strategy specification} to allow selection of the strategy to use.

    In this context, our paper makes the following contributions:
    \begin{enumerate}
        \item we consider different types of combinatorial reoptimisation problems in the context of \ac{MDE}  and identify the types of contextual changes that can affect optimisation problems (see \cref{sec:problem_types});
        \item we identify four types of solution strategies for increasingly more powerful (and more expensive) reoptimisation problem specifications, based on the type of contextual change (see \cref{sec:strategies});
        \item we demonstrate feasibility of our approach in a motivating case study about allocating \acp{TA} to teaching sessions (see Sections~\ref{sec:motivating_example} and \ref{sec:implementation});
        \item we then move towards the automatic generation of the reoptimisation problem specifications and solver strategies, and formulate future research questions in \cref{sec:roadmap}.
    \end{enumerate}

%% file: motivating_example.tex

\section{Motivating example}
\label{sec:motivating_example}

	\insertFigure[targetwidth=0.75\linewidth,%
					caption={Simplified teaching assistant scenario metamodel.}%
					]%
					{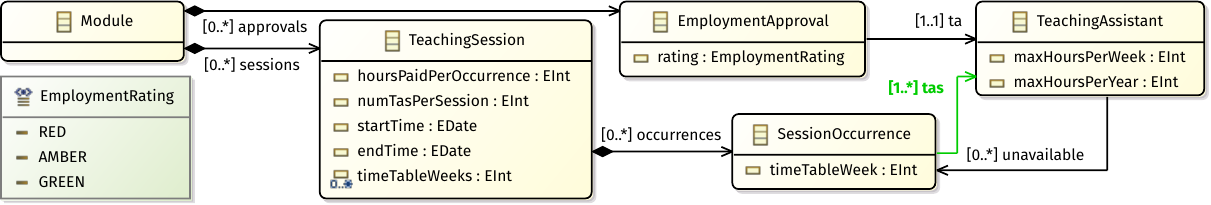}

	To illustrate our vision, we have chosen the assignment of teaching assistants (\acp{TA}) to university courses~\cite{Qu+17}.
	Our version of this scenario
	is based on a real-world problem present at the Department of Informatics at King's College London, where the goal is to determine a valid assignment plan for several courses and the \acp{TA} in a given semester.

	\Cref{figure:metamodel} shows a simplified version of the metamodel that we developed to capture the conditions of the original problem.
	In the scenario, courses are modeled as \texttt{Module}s, whereby each \texttt{Module} can have multiple (possibly recurring) events that we call \texttt{Teaching\-Session}s.
	An example instance could be the course \texttt{Mathematics I} with multiple weekly exercise groups, each of which is an individual \texttt{Teaching\-Session}.
	Every \texttt{Teaching\-Session} can take place in multiple \texttt{Week}s, and for every occurrence, there must be exactly one \texttt{Session\-Occurrence}.
	This type is used to assign a set of \texttt{TeachingAssistant}s to the \texttt{Teaching\-Session}s in each week.
	The number of necessary \acp{TA} per actual teaching session is defined by the attribute \texttt{numTasPerSession}.
	This design allows assigning multiple \acp{TA} to different session occurrences. For example, it can be used to model a \ac{TA} substitution for a particular date only, while also allowing, e.g., bi-weekly meetings.
	For the selection of \acp{TA} according to their skill, the metamodel contains the type \texttt{Employment\-Approval} in which a \ac{TA} can be marked as \texttt{GREEN} (best fit), \texttt{AMBER} (possible fit but not ideal), or \texttt{RED} (the \ac{TA} cannot supervise the session).
	The design decisions in this metamodel are implicitly based on the real-world planning system of the department of informatics at King's College London.
	However, other ways of modeling the same scenario are possible.

	The goal is to find an assignment from \texttt{Teaching\-As\-sis\-tant}s to \texttt{Session\-Occurrence}s, and store it as \texttt{tas} edges in the model.
	These edges determine which \texttt{Teaching\-Assistant} is used to supervise an instance of a \texttt{Teaching\-Session}.
	As a result, the creation of each assignment edge is a variable in the optimisation problem to be solved, meaning that all possible assignment edges define the complete search space of the optimisation.
	A feasible solution ensures that all constraints are fulfilled, for example, the weekly work time limitation or the time limitation for a whole semester of each \ac{TA}.
	Since multiple feasible solutions usually exist for a given problem instance, we also aim to optimise the \texttt{Employment\-Approval} ratings of the \acp{TA} for their assigned courses.
	The objective function for the scenario, therefore, consists of the possible assignments of \acp{TA} to \texttt{Session\-Occurrence}s, with each possible assignment having a weight corresponding to the \ac{TA}'s approval.
	The overall goal is to find an assignment that utilises the best possible pairing of \acp{TA} and sessions.

	We use the \ac{GIPSL}~\cite{gipsGCM2022} to specify the \textit{original} problem from multiple components:
	Firstly, graph patterns and \ac{GT} rules are original search operators that can be used to inspect structural model properties and transform the model.
	%
	%
	We specified a single \ac{GT} rule \texttt{assign\-Ta} shown in \cref{figure:assign-ta-gt-rule} creating an assignment edge, as described above.
	In addition to the structural requirements, the \ac{GT} rule also includes an attribute condition.
	This condition only allows for \acp{TA} that have at least the approval status \texttt{AMBER}.
	If the rule is applied, an edge \texttt{tas} (shown in green and with the label \texttt{++}) will be created between \texttt{occurrence} and \texttt{ta}.
	This edge represents the assignment of the \texttt{Teaching\-Assistant} \texttt{ta} to the \texttt{Session\-Occurrence} \texttt{occurrence} and maps to a decision variable.
	%
	%
	The next component of our original problem specification consists of \ac{GIPSL} \textit{constraints} restricting the solution space of the optimisation problem.
	For example, the specification contains a constraint to ensure a feasible solution assigns exactly the amount of \acp{TA} to a \texttt{Session\-Occurrence} required by the respective \texttt{Teaching\-Session}.
	%
	%
	Lastly, we define an objective function to encourage the solver to select ``green'' \acp{TA} when possible and ``amber'' \acp{TA} when necessary.
	Consequently, the objective function contains all the possible assignment variables and gives the ``green'' \acp{TA} a higher weighting than the ``amber'' \acp{TA}.
	Based on the complete \ac{GIPSL} specification\footnote{All related \ac{GIPSL} projects including some problem instances, a problem generator, and multiple solution implementations can be found in the repository https://github.com/Echtzeitsysteme/reoptimisation-paper-2025-example.}, the \ac{GIPS} tool generates a \ac{MILP} specification to optimise all possible \ac{TA} assignment models that conform to the metamodel shown in \cref{figure:metamodel}.
	The \ac{GIPS} tool is executed to apply a single or multiple \ac{GT} rule(s) selected by the \ac{MILP} solver.
	Since \ac{GIPS} uses high-level specifications as input to generate the \ac{MILP} problem generator automatically, we can easily use it to implement the motivating example.
	Compared to a plain \ac{MILP} approach, we can adapt the high-level specification and must not deal with the generation of equations ourselves.

	\insertFigure[targetwidth=0.8\linewidth,%
					caption={\ac{GT} rule \texttt{assign\-Ta} that can be used to assign a \ac{TA} to a specific session occurrence. This assignment is modelled by creating a new edge (\texttt{++\,tas}) between \texttt{occurrence} and \texttt{ta}. All context information is visualised in black and white.},%
					onecolumn=true]%
					{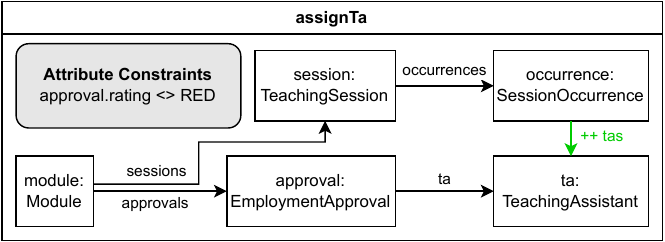}

%% file: problem_types.tex

\section{Specifying changes and reoptimisation problems}
\label{sec:problem_types}

	In this section, we provide an overview of the possible changes to an optimisation problem and discuss what this means for the \emph{specification} of the resulting reoptimisation problem.
	The following section sketches possible procedures for solving these problems.

	As outlined in the introduction and our example, an instance of an optimisation problem is a model, where a metamodel, together with constraints and model transformations, defines the search space. The quality of the solutions is measured by objective functions.
	In our example, the following are instances of realistic changes that can happen after a solution to an original optimisation problem has been computed:
	\begin{enumerate}[wide, labelwidth=!, labelindent=0pt, label={\emph{Scenario \arabic*:}}]
		\item \emph{Unavailability for assigned session.} A \ac{TA} gets blocked for a specific time frame to which they previously had been assigned.
		\item \emph{Reduction of working hours.} A \ac{TA} has reduced their weekly working hours to the extent that they can no longer deliver several sessions.
		\item \emph{Complex set of non-availabilities.} Multiple \acp{TA} get blocked for individual dates, and there is no simple solution to reassign them, e.g., by swapping them one by one.
		\item \emph{Vacuous unavailability.} A \ac{TA} gets blocked for a specific time frame for which they are not assigned.
	\end{enumerate}
	Thus, to support users in the construction of reoptimisation problems, we need a language that allows to express (sets of) specific changes.
	This \emph{Problem Change Specification} needs to enable expressing changes to (almost) all ingredients of the specification of the optimisation problem and its solutions.
	Already in the small set of examples above, the changes in Scenarios~1, 3, and 4 are changes to the model (either problem instance or solution), namely, edges of type \texttt{unavailable} are introduced.
	The change in Scenario~2, however, in principle could concern any ingredient of the specification:
	\texttt{maxHoursPerWeek} can be hard-coded in the metamodel (e.g., because of legal requirements), defined in the instance model, given as an additional constraint, or realised as an application condition in \texttt{assign\-Ta}.
	In our motivating example in Section~\ref{sec:motivating_example}, we set individual \texttt{maxHoursPerWeek} per \ac{TA} in instance models and add a constraint that forbids weekly assignments to surpass that number.
	Thus, while one might argue that changes to the objective functions or changes beyond attribute values to the metamodel are rather redefinitions of a problem and should not be considered as reoptimisation, the \emph{Problem Change Specification} definitely needs to allow for expressing complex changes to the different ingredients of an optimisation problem.

	Given an original problem, its solution, and the description of a specific change, we need to \emph{specify the resulting reoptimisation problem}.
	Again, this can affect all ingredients of the specification of the original optimisation problem; we discuss the effects on search operators and the objective function(s) in the next section and here focus on constraints and metamodel.
	First, and obviously, we have to integrate the specific change, e.g., replace one of the original constraints by an edited one or optimise an edited problem instance.
	More intricate, however, are the indirect modifications that typically arise.
	\begin{description}[wide, labelwidth=!, labelindent=0pt]
		\item[Constraints] Regularly, there are parts of the original solution that cannot or should not be changed.
			This often pertains to time (e.g., assignments of \acp{TA} in the past cannot and assignments in the very near future should not be changed).
			Additionally, a user may wish to fix certain parts of the previous solution explicitly.
			In our example, there might be a professor who is particularly averse to changing their \acp{TA} mid-term, so one decides not to change the assignments of those \acp{TA} (Instead of using constraints, this could also be modeled by assigning high costs to certain changes in the objective function).
		\item[Metamodel] The metamodel might not be defined on the level of granularity that allows marking elements as belonging to an original solution or to express temporal properties like being in the past or future.
			Thus, to be able to express desirable constraints, it might be necessary to first amend the metamodel.
	\end{description}

%% file: strategies.tex

\section{Specifying solution strategies for reoptimisation}
\label{sec:strategies}

	Different changes to an optimisation problem can have different effects on the original solutions.
	Therefore, it is desirable to have different \emph{strategies} available to react to these changes.
	In our example, Scenario~1 renders the original solution invalid, but we would expect a valid and at least locally optimal solution to exist nearby (e.g., one that can be reached by swapping two assignments).
	In Scenario~2, the original solution again becomes invalid, but now several sessions of the respective \ac{TA} have to be reassigned.
	This still provides \enquote{locations} that need repair (namely, the sessions of that \ac{TA}), but typically leads to a combinatorial explosion of repair possibilities.
	Similarly, Scenario~3 renders the original solution invalid and makes local repair difficult.
	Scenario~4 neither affects the validity nor the optimality of an original solution.

	Furthermore, it may be preferable to present users not only with a new solution but also with an explanation of the differences between the original and new solutions, or even a script that produces the new solution from the original one.

	In response to these challenges, we propose two fundamental types of strategies: ``plaster'' rules---which enable a local search for a new solution---and full recomputation.

	\subsection{Local-neighbourhood search with ``plaster'' rules}

	Here, we aim to fix the problems introduced by a change through a localised update to the original solution.
	This is comparable to local neighbourhood approaches in the existing literature (e.g.,~\cite{MV11,VPHK12}).
	The key idea is that we replace the original search operators with new operators.
	These new operators encode the changes needed to transform an original solution into a new solution---we call them ``(sticking) plaster rules'' as they provide a localised fix for a small problem.
	We define a local neighbourhood for the search by ensuring the new operators are only applicable to parts of the original solution that contribute to its invalidity.
	Additionally, we may include constraints or objective functions that restrict the number of times a search operator can be applied.
	Plaster-rule strategies are naturally good at producing edit scripts.

	We can identify three, increasingly more complex, plaster-rule strategies, though others may be possible:

	\subsubsection{``Basic'' plasters}

		These solve a single constraint violation in the most direct way possible.
		They provide a localised fix and, in particular, avoid changing any existing edges, nodes, or attributes not directly involved in the constraint violation.
		A single plaster rule must be applied exactly once.

		Basic plaster rules should be strongly consistency-im\-prov\-ing rules~\cite{Kosiol+21}; that is, any application should fix all constraint violations, leaving the optimiser to pick a local optimum, i.e., the best fitting application that restores consistency.
		This can, for example, be achieved by systematically adding appropriate application conditions to the original search operators so they can only be applied where a constraint has been violated.

		In the \ac{TA} example, the basic plaster rule \texttt{assign\-Ta'} was derived from the original rule \texttt{assign\-Ta} to search for a \texttt{Session\-Occurrence} without a \texttt{Teaching\-Assistant} (because they are not available anymore).
		The new rule will assign another available \ac{TA} and, hence, repair the original solution.
		All other assignments will be kept the same.

	\subsubsection{``Smart'' plasters}

		These are like basic plasters but may selectively change parts of the model not directly affected by the constraint violation.
		Thus, they may solve situations where no basic plaster is available, possibly resulting in a better overall solution.
		Again, exactly one smart plaster must be applied, and it must be strongly consistency improving.

		Swapping the allocation of two \acp{TA} is an example of a smart plaster.
		This may work even when there is no \ac{TA} that still has sufficient time available to take on an additional session occurrence; the basic plaster above would fail in this case.
		In our example implementation, this is achieved by creating a \ac{GT} rule \texttt{swap\-Tas} to swap two \acp{TA}.
		For this smart plaster, there are additional constraints that only allow the swapping of unavailable \acp{TA} with available ones.
		Hence, the repair depends on the fact that for each blocked \ac{TA} \texttt{t1}, there is at least one \ac{TA} \texttt{t2} that can take on the blocked session and vice versa.

	\subsubsection{Plaster sets}
		Some problem types do not lead to an easily identifiable constraint violation in a single location.
		For example, in Scenario~2 from \cref{sec:problem_types}, we may need to reallocate several of sessions if there is no single session the \ac{TA} could give up to bring them below their new weekly hours cap.
		However, it would potentially still be beneficial to apply specific plaster rules limited to only the allocations of the ``problematic'' \ac{TA}.

		A `plaster set' strategy generates one or more basic or smart plaster rules, but does not limit them to exactly one application.
		Each application of a plaster rule is still expected to improve the consistency of the resulting solution, though a single application may not completely fix any constraint violation.
		Lauer et al.~\cite{Lauer+24} introduce c-increasing rules, which capture this idea of partially improving the consistency of a graph-based model.
		Rules will typically be defined such that their matches all include solution elements that contribute to the problem to be solved.
		In the \ac{TA} allocation example, this might involve defining rules such that the affected \ac{TA} is always part of the rule match to avoid a complete recalculation.
		Since plaster sets only allow changes of the model in the neighbourhood of a \ac{TA} with violated constraints, it searches for a local optimum.
		In comparison to a normal plaster, this allows a specific plaster \ac{GT} rule to be executed multiple times---in our example, the previously described basic plaster \ac{GT} rule \texttt{assign\-Ta'} without the requirement of the \ac{TA} being ``blocked''.
		Since multiple rule applications are allowed, the solver can choose multiple available \acp{TA} to be the substitute for a \ac{TA} whose time limit is exceeded by the original solution.

	\subsection{Full recomputation}

		Some problems are too complex to be solved with a localised `plaster' and in some cases, a local optimum may not be enough.
		For these cases, we can completely recompute a new (globally) optimal solution.
		However, we need to amend the original problem specification to ensure `minimal perturbation'~\cite{Bartak+04} or `least change'~\cite{MJC16}.
		Specifically, we must:
		\begin{enumerate}[wide, labelwidth=!, labelindent=0pt]
			\item Adapt the metamodel by generating parallel `shadow' structures for any modified element as part of the optimisation.
				These are used to keep track of the original solution.
			\item Add an additional term to the objective function (or a separate objective function for multi-objective problems) that uses the shadow structures for computing the difference between the original and the new solution, penalising any delta.
		\end{enumerate}

		Because we produce a completely new solution, we need an additional processing step to derive sequences of adaptation operations.
		However, in many cases, we may be able to reuse the shadow structures we have generated to derive the delta and, from this, reverse engineer a possible sequence of adaptation operations.
		In the example implementation, we added a duplicate assignment edge \texttt{previousSolutionTas} within the metamodel.
		When running the complete recomputation of the problem, the program uses this new edge type to capture the original computed assignment of \acp{TA}.
		All \ac{GT} rules and constraints of the reoptimisation specification are the same as in the original specification.
		The only change lies within the objective function, where we add a penalty if a previously assigned edge will not be chosen again.
		This ensures the solver can change the whole assignment if necessary, but since it wants to minimize the costs, the new solution is as close to the original solution as possible.

%% file: implementation.tex

\section{Proof of Concept Implementation}
\label{sec:implementation}

    So far, we presented the motivating example (\cref{sec:motivating_example}), different types of reoptimisation problems that could occur (\cref{sec:problem_types}), and a selection of search operators to solve the problems (\cref{sec:strategies}).
    This section aims to investigate the feasibility of our approach by applying the reoptimisation approach described above to the example presented in \cref{sec:motivating_example}.
    For this, we developed multiple \ac{GIPS}-based solutions that represent the different strategies presented in \cref{sec:strategies}.
    We use \ac{GIPS} implementations of Scenario~1--3 from \cref{sec:problem_types} to answer the following \acp{RQ}:
    \begin{enumerate}[label={\textbf{RQ\arabic*}}, wide, labelindent=0pt]
        \item What strategies can be used to solve the reoptimisation problems resulting from different change types?
        \item How close is the solution of the reoptimisation problem to the solution of the original problem?
    \end{enumerate}
    For the three experiments, we used synthetic models generated from up to $7$\,modules, $20$\,\acp{TA}, and a planning horizon of $4$\,weeks.
    We did not consider any timing aspects from the past, i.e., the implementations could not be configured to keep past \ac{TA} assignments fixed.

    Addressing \textbf{RQ1,} \cref{tab:eval-identical-assignments} shows the results of applying our four strategies to Scenarios~1--3.
    Every entry without a dash indicates that the particular strategy can be used to find a solution.
    All strategies successfully solved the most basic scenario (Scenario~1), which involves only one blocked \ac{TA} and no other modifications.
    This is because it is possible to solve Scenario~1 by (re)allocating a single \ac{TA}, which all implementations support.
    Scenario~2, in which the weekly working hours of a \ac{TA} are reduced, can only be solved using a ``plaster set'' or by fully recomputing the optimisation problem.
    This makes sense, as there is no single location that represents the constraint violation, but multiple reallocations of \acp{TA} are required to fix the constraint violation.
    Ultimately, the only strategy to solve the most challenging Scenario~3 is to fully recompute all assignments.
    This is because, in this scenario, there is no single \ac{TA} available to replace the blocked \ac{TA}, nor are there any other sufficiently approved \acp{TA} available for the specific timeframe in which the respective \ac{TA} is blocked.
    Therefore, it is evident that our implementations of the simpler strategies are insufficient to solve this scenario.

    \begin{table}[tbp]
        \caption{Number of identical assignments chosen by the different reopt. strategies out of the total number of assignments. A dash denotes that the strategy cannot solve a scenario.}
        \centering
        \begin{tabular}{lccc}
            \toprule
            \textbf{Strategy} & \textbf{Scenario~1} & \textbf{Scenario~2} & \textbf{Scenario~3} \\
            \midrule
            Basic plaster & $33/34$ & - & - \\
            Smart plaster & $32/34$ & - & - \\
            Plaster set & $33/34$ & $30/34$ & - \\
            Full recomputation & $33/34$ & $30/34$ & $0/34$ \\
            \bottomrule
        \end{tabular}
        \label{tab:eval-identical-assignments}
    \end{table}

    Regarding \textbf{RQ2}, we observed the number of identical assignments, i.e., the assignments chosen by the reoptimisation solver that were equal to the original solution, which can be seen in \cref{tab:eval-identical-assignments}.
    In Scenario~1, all approaches but the ``smart plaster'' achieved $33/34$ identical assignments.
    This makes sense since the swap rule must always change two previously assigned \acp{TA} to repair a constraint.
    Regarding Scenario~2, it can be observed that the ``plaster set'' and the full recomputation were able to fix the violation by reallocating three sessions of the overloaded \ac{TA} to other \acp{TA}.
    Only the implementation to fully recompute all assignments was able to solve Scenario~3, which was intentionally designed such that none of the assignments of the original solution could be kept.

%% file: related_work.tex

\section{Related work}
\label{sec:related_work}

	Various aspects of combinatorial reoptimisation have been considered in the literature.
	We will relate our research objectives to closely related work in this field.
	We will also consider approaches to graph and model repair, since the models used to specify the problem instances are subject to incremental changes.
	Graph and model repair processes should typically entail a specific strategy, the least possible change.

	Various theoretical approaches to reoptimisation have been developed to solve different types of optimisation problems, including combinatorial problems (e.g. \cite{SST+18}), dynamic graph problems (e.g. \cite{EGI+99}), and computationally hard problems (e.g. \cite{EAB+11}).
	The general objective of these studies is to efficiently compute an optimal solution for the modified problem instance.
	If there is no cost associated with transitioning between solutions, the resulting solution may differ considerably from the original one.
	However, in \cite{SST+18}, the minimisation of transition costs is also a subject of interest.
	The authors consider reoptimisation problems that involve two challenges: (1) computing a (close to) optimal solution for the new problem instance, and (2) efficiently converting the current solution to the new one.
	They present theoretical results for combinatorial reoptimisation, including objective functions that address the above challenges simultaneously.
	The authors show that reoptimisation involving transition costs may be harder than solving the underlying optimisation problem.
	In this sense, the original solution plays a restrictive role rather than helping to solve the modified problem instance.
	This challenge is similar to incremental optimisation (see e.g. \cite{COG+18}), in which a partial solution is optimised step by step.

	The reoptimisation problems that we consider are also closely related to minimal perturbation problems~\cite{Bartak+04}, each of which consists of a \ac{CSP}, an original (partial) solution, and a distance function between solutions.
	A solution to a minimal perturbation problem is a solution to its \ac{CSP}, for which the distance to the original solution is minimal.
	They arise for all kinds of scheduling problems, e.g., university timetabling when changes occur to staff employment (e.g., \cite{PWE+17}).
	{\em While those papers focus on the complexity of these problems, they do not aim to automatically generate the specifications of the reoptimisation problems or solver strategies from the original ones.}

	Model repair is required when a model changes in such a way that it becomes inconsistent with given constraints.
	Various approaches to model repair exist; a feature-based classification of these approaches up to 2016 is provided in~\cite{MJC16}.
	The type of problems and solver strategies considered in this paper are similar to those in rule-based model repair~\cite{BMM+09,TOL+17,OPK+21,MKA+23,Lauer+24}.
	{\em Although model repair can be considered as an optimisation problem -- for example, making the least possible change to the model --  it is not an approach to solving optimisation problems or even reoptimisation problems.}

%% file: roadmap.tex

\section{Future Research Questions}
\label{sec:roadmap}

	So far, we have (i)~presented the problem of reoptimisation in the context of \ac{MDE}, (ii)~sketched some potential solution strategies, and (iii)~reported on a manual implementation of one concrete combinatorial optimisation problem, namely the \ac{TA} allocation.
	\emph{In the long run, our goal is to use \ac{MDE} techniques to automate the specification and solving of reoptimisation problems, given an original optimisation problem, a solution to it, and a description of changes.} 
	Our considerations in this paper lead to the following \acp{FRQ}, the answers to which enable the realisation of our vision.

	\begin{enumerate}[label={\emph{FRQ\arabic*.}},left=0pt, wide, labelwidth=!, labelindent=0pt]
		\item \emph{How can a good \ac{DSL} for \emph{Problem Change Type Specifications} look like?}
				Designing such a language can be based on established principles of language engineering~\cite{Laemmel18, WasowskiB23}.

		\item \emph{What types of reoptimisation problems exist?}
				In \cref{sec:problem_types}, we discuss the different problem parts that can change to trigger a reoptimisation step.
				To further develop our research vision, we will need to make this catalogue more systematic, exploring the different changes that can happen to a model and how they inform the reoptimisation problem that needs to be generated.
				Developing a syntactic categorisation based on the action by which and the ingredient that was changed (like ``attribute value change in metamodel'', ``deletion of element in instance model'', ``strengthening of constraint'') is probably straightforward.
				The more interesting question is how such changes affect the reoptimisation problem that needs to be generated (in particular, the effect on the validity and optimality of original solutions and the neighbourhood in which we can expect to find good new solutions) and whether we can link syntactic types of changes to their effects.
				To explore potential links, we want to systematically scan the existing theoretical literature on reoptimisation.
				Moreover, there exist works on model evolution \cite{DamG14} or model repair~\cite{TOL+17, Nassar2020}, where catalogues of possible model changes are linked to appropriate responses that might prove useful.

		\item \emph{What is a complete set of strategies for solving reoptimisation problems and how can they be constructed automatically?}
				In \cref{sec:strategies}, we list four strategies for problem generation.
				Is this list complete or are there other strategies that can be employed in certain cases?
				Regarding the construction of plaster rules for the strategies we suggest, there is a wealth of literature to build upon, including literature on model editing~\cite{BMM+09, Tinnes23}, (rule-based) graph and model repair~\cite{MJC16, Nassar2020, MKA+23, Lauer+24}, the composition of \ac{GT} rules~\cite{KosiolT23, Kreowski25}, and rendering rules to make them validity-preserving or even validity-improving~\cite{Nassar2020, Burdusel+21, Kosiol+21, Horcas+22}.
				While especially smart plasters will depend on the specific problem and addressed change, this research provides us with many promising options to explore for the automation of the construction of (smart) plaster rules.

		\item \emph{How can we automate the derivation of a reoptimisation problem specification from the original optimization problem specification, its solution, and a description of the changes to the problem?}
				In general, generating the reoptimisation problem should be encoded as a model-to-model transformation.
				While generating new search operators or adapting the objective functions seems feasible, automating suitable metamodel refinements and automatically choosing a suitable solution strategy pose serious challenges.
				Some user interaction (e.g., deciding whether a globally or locally optimal solution is needed) is probably unavoidable.

		\item \emph{How does automated generation of (re-)optimisation problems affect solver efficiency?}
				What are the performance impacts of different strategies? How do our strategies mesh with typical strategies for making large optimisation problems tractable, such as decomposition and relaxation?
				We can potentially build on literature from constraint-solving and optimisation communities---for example, work on \textsc{Conjure}~\cite{Akguen+22,Stone+24}.
	\end{enumerate}

	We plan to further explore these research questions building on case studies from resource allocation, healthcare scheduling, and others.
	We believe this will enable us to make significant contributions to the operations-research community and beyond.
	Currently, our vision is explicitly limited to combinatorial problems, of which there are many.
	Studying whether similar ideas could be applied to non-combinatorial optimisation problems is an open question.